# Genetically optimized All-dielectric metasurfaces

VICTOR EGOROV, MICHAL EITAN AND JACOB SCHEUER*

*School of Electrical Engineering Tel Aviv University, Ramat Aviv, Tel-Aviv 69978, Israel*
*\*kobys@eng.tau.ac.il*

**Abstract:** We present and study theoretically a new design approach for obtaining wide angle, highly efficient, all-dielectric metasurfaces. As a concrete example we focus on optimizing flat beam deflector for both the infra-red and visible spectral regions. Transmission efficiencies of up to 87.2% are obtained theoretically for deflection angle of 65° in visible (580nm) spectrum and up to 82% for deflection angle of 30.5° at telecom wavelength (1550nm). The enhanced efficiencies at wide deflection angles are obtained by genetic optimization of the nano-structures comprising the metasurface. Compared to previously employed design approaches, our approach enhances the transmission efficiency substantially without sacrificing rectangular grid arrangement and facilitates the realization of wide angle flat deflectors and holograms/lenses.

## 1. Introduction

During the last few years much research efforts have been focused on designing flat metasurfaces facilitating subwavelength light control and flat optical components such as lenses, holograms and more. Plasmonic based single-layer metasurfaces suffer from the intrinsic high losses of metals and very low transmission efficiency at optical frequencies [1, 2]. All-dielectric metamaterials, on other hand, exhibit low absorption losses in the infrared (IR) and visible spectral ranges, rendering them attractive for developing flat diffractive optical components. In order to achieve high transmission efficiencies many dielectric metasurfaces implement the Huygens surface principle, by utilizing an overlap of electric and magnetic dipolar Mie-type resonances of the constituent high refractive index elements to reduce significantly backscattering of the impinging light. Consequently, the complete $2\pi$ phase shift coverage required for metasurfaces can be obtained with very high transmission efficiency over relatively broad wavelength ranges [3, 4]. There are two common design techniques for flat all-dielectric metasurfaces: the first utilizes geometric or Pancharatnam-Berry (PB) phase and the second are based on tailoring the physical dimensions of the single elements in the structure to obtain the desired phase response.

The Eigen states of PB phase are left and right circularly polarized beams. Local phase spanning of $0...2\pi$ can be obtained by spatially rotating the antenna (a single element in the array) orientations [5]. Indeed, high efficiency devices such that holograms and lenses at the visible and IR bands, based on geometric phase design, have been recently demonstrated [6-11]. However, PB phase approach is inherently limited to circularly polarized light and is unsuitable for linear polarization or for polarization independent metasurfaces.

The second design technique utilizes the physical dimensions of the elements comprising the metasurface in order to obtain the desired local phase retardation of the transmitted wave [12-20]. Square (rectangular) [16] and circular (elliptic) [17, 18] dielectric nano antennas are frequently employed for designing and realizing such metasurfaces. For example, beam deflecting metasurfaces designed using this approach have been shown to achieve high transmission efficiencies of nearly 68% at wide angles of up to 65° [13, 17, 19] in visible spectrum (580nm). The Eigen states of this design approach are linearly polarized along long and short axes of the individual elements or, alternatively, linear polarizations corresponding to the axes of the lattice [19]. Complete $2\pi$ phase span has also been demonstrated using this approach [3, 17, 18]. Since the nano antenna elements are not required to be anisotropic and can be of circular or square shape, this approach facilitates the realization of polarization independent devices [19]. Here we focus on metasurfaces design based on the second approach.

The great potential of metasurfaces is their ability to induce arbitrary phase profiles on an impinging beam. The actual realization of such metasurface requires the construction of a library of nano-structures with known transmission properties (phase, amplitude, etc.). These building blocks constitute the "pixels" of the metasurface and provide the desired phase (and amplitude) response. The conventional approach for obtaining these properties is based on simulating infinite periodic arrays of single antennas with varying dimensions. The obtained response of phase/amplitude is then associated with the individual unit cell [3, 4, 16, 17]. For example, a metasurface can be constructed by an array of circular disks with fixed thickness and period while changing the radii of the disks in order to obtain varying phase shifts. A commonly used structure for benchmarking metasurfaces design strategy is a beam deflector [17, 21, 22]. In addition to being an important component with various possible applications, the performance of the beam deflector (primarily its transmission efficiency) can help to adequately compare different designs approaches. Furthermore, the building blocks of such optimized beam deflector are highly useful for realizing highly-efficient arbitrary holograms to be projected at the same angle the deflector is designed for [21, 23].

In this letter we present a new approach for designing highly efficient wide angle flat deflectors at telecom and visible wavelengths, by utilizing dielectric metasurfaces. Note that we focus of the beam deflection application because it facilitates simple comparison to other metasurfaces design approaches. Although highly efficient beam deflection designed by specific approaches (see e.g. [12, 13]) have been demonstrated, an efficient metasurface deflector design can be extended for designing more complex flat components such as lenses and holograms [12, 13, 18, 20, 23, 24]. The new approach, based on genetic algorithm optimization, exhibits superior efficiency compared to the more commonly employed design approaches reported previously [3, 13, 17, 21].

## 2. Design and comparison

Figure 1(a) illustrates a schematic of the structure of the proposed device, consisting of Si disks with fixed thickness (280nm) and varying radii, arranged in a square lattice. The disks are positioned on top of 500nm thick $SiO_2$ pedestals which are part of the etched substrate. The periodicity of the array is 764nm in both transverse directions. Decker et al. [3] noted that it is the refractive index contrast between the surrounding medium and nano disks which primarily determines the modal confinement of resonance and that appropriate choice of the clad material can improve substantially the transmittance performance. For the device studied here, we choose SU-8 photoresist as the clad material. The refractive index of SU-8 at 1550nm is 1.574 which is close to desired refractive index of 1.66-1.68 [3]. Both the Si disks and the etched substrate are covered with the upper cladding layer (SU-8 resist). The upper cladding is covered by an additional layer which serves as an anti-reflective (AR) coating whose properties and impact on the device performances are further discussed below.

First, we study deflectors without the AR layer in order to optimize the geometry of the metasurface. We assume that the structure is illuminated by a linearly polarized plane wave (ŷ in Fig. 1(a)), propagating in the positive ẑ direction. The metasurface is designed to deflect the beam in the x direction. At $\lambda_0$=1550nm the refractive indices of the constituent materials are approximately: $n_{Si}$=3.48, $n_{SiO2}$=1.44, $n_{SU-8}$=1.574 with relatively small absorption. In order to obtain beam deflection, a linear gradient phase response is needed according to the generalized Snell law [25, 26]:

$$\sin(\theta_{trn})n_{trn} - \sin(\theta_{inc})n_{inc} = \frac{\lambda_0}{2\pi}\frac{d\phi}{dx} \quad (1)$$

where $\theta_{inc}(\theta_{trn})$ and $n_{inc}(n_{trn})$ are respectively the wave propagation angle with respect to the normal to the interface and the refractive index in the incidence(transmittance) regions. Note that this description, which is based on the modified Snell law, is equivalent to the beam steering criteria often used in phased array antennas [27]. For normal incidence, we consider a constant phase gradient along the x direction by linearly distributing $2\pi$ phase shift of the length of the deflector super cell - $\Lambda$. The propagation angle of the transmitted wave is then given by:

$$\theta_{trn} = \sin^{-1}\left(\frac{\lambda_0}{n_{trn}\Lambda}\right) \quad (2)$$

And again, note that Eq. (2) corresponds to the steering angle of an array consisting of equally spaced (*d*) and phase-shifted ($\Delta\phi$) emitters such that $\Lambda=2\pi d/\Delta\phi$. It should be also noted that the metasurface based deflector is a *periodic* structure with a period corresponding to the length of the supercell. Referring to Eq. (2), it can be seen that the desired deflection angle corresponds to the *first Bragg diffraction order* of the metasurface. Moreover, as the length of the supercell is larger than the wavelength, several Bragg diffraction orders may exist.

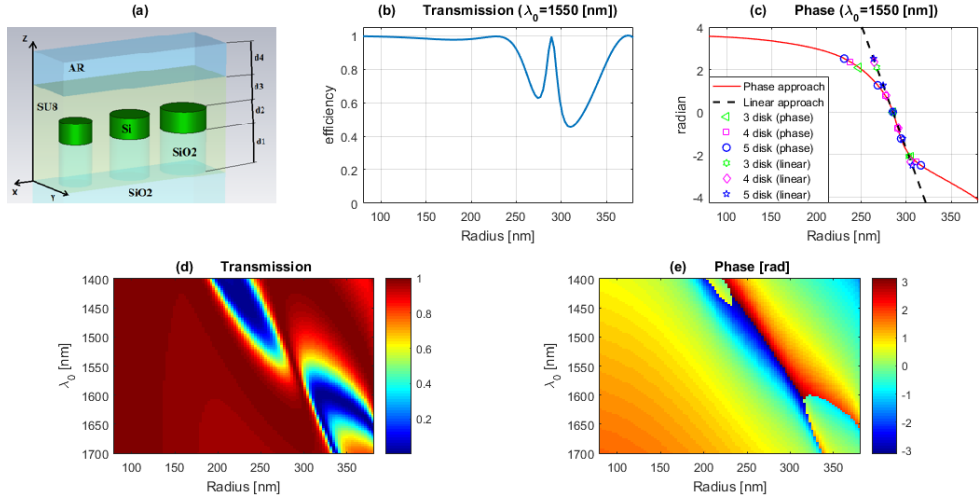

**Fig. 1. (a)** Schematic of a 3-disk super-cell deflector, d1 – $SiO_2$ etch depth, d2 – Si disk thickness, d3 – upper clad thickness, d4 – AR layer thickness. **(b)** Transmission intensity of periodic array at 1550nm for varying disk radius. **(c)** Transmittance phase at1550nm. **(d)** Numerically calculated transmittance intensity and **(e)** phase maps of periodic arrays of disks as a function of the disks radius.

Figure 1 depicts the amplitude (Figs. 1(b) and 1(d)) and phase (Figs. 1(c) and 1(e)) spectral response of the transmitted wave through a periodic array comprising identical, equally spaced, disks as a function of their radius. Recall that the spacing between adjacent disks is 764nm in both transverse directions. The spectral response was calculated by the rigorous coupled wave analysis (RCWA) approach [28] and verified using commercial FDTD software [29]. The overlap of the electric and magnetic resonances can be clearly identified by the nearly unity transmission in a narrow spectral window around $\lambda_0$=1550nm where the backward scattering is strongly suppressed. Note that the transmission efficiency can be improved by increasing the periodicity but at the expense of reducing the deflection angle (for a fixed number of disks per supercell).

As mentioned above, the common approach for constructing a deflecting metasurface is based on utilizing scatterers providing linear, equally spaced, phase increments based on maps such as shown in Fig. 1 [3, 17]. Note that as the phase span of the curve depicted in Fig. 1(c) is larger than $2\pi$, the choice of radii is not unique. There is also no guarantee that all possible choices provide similar performances. Another, even simpler design approach, which employs disks with linearly increasing radii, has been presented in [22].

To illustrate and compare the effectiveness of these approaches we constructed metasurface based deflectors consisting of 3, 4 and 5 disks per unit-cell, corresponding to free space deflection angles of 42.6°, 30.5°, and 24° respectively. The choices of radii of the phase-map approach for each design are marked on the phase line in Fig. 1(d). For the linear disk radii approach, we draw a straight line intersecting the resonant overlap point with the slope of the phase curve at the same point (dashed black line in Fig. 1(c)). The choices of the disks radii of this approach are also shown in Fig. 1(c).

Figure 2 depicts simulation results of the transmitted diffraction efficiencies (TDE) at the designated angles (42.6°, 30.5°, and 24° respectively) for the phase-map and the linear radii design approaches as well as for an optimized approach which is further discussed below. We define the Transmission diffraction efficiency as the fraction of the total incident power which is deflected to the designed angle. The phase-map design approach provides reasonable transmission efficiency of approximately 50% at small deflection angles (Fig. 2(c)), which decreases substantially at larger angles (less than 25% for 42.6°). We note that the obtained deflection efficiencies are substantially lower than the transmission efficiencies found for the infinite periodic arrays (Fig. 1(e)). We attribute this difference to coupling effects between adjacent disks which modify the actual phase response of the individual disks when placed in a non-periodic arrangement. We also note that although such coupling effects have been shown to have substantial impact in plasmonic metasurfaces [21], they are often effectively neglected when designing dielectric metasurfaces with an arbitrary phase response [15].The impact of coupling effects in such metasurfaces was also demonstrated by Chong et al. [30], who used such effects in order to control the phase response of dielectric Huygens metasurfaces.

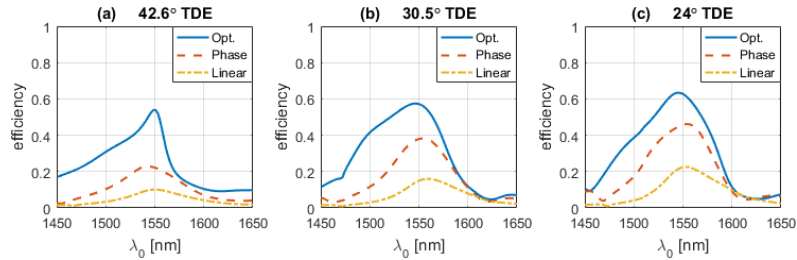

**Fig. 2.** Spectral efficiencies (without AR coating) of the deflectors designed to **(a)** 42.6°; **(b)** 30.5°; **(c)** 24°.

*2.1 Genetic algorithm optimization*

Thus, it seems that the performances of dielectric metasurfaces can be further improved by properly optimizing the design and parameters of the disks. Due to the relatively large number of available parameters, an exhaustive scan over the complete parameter space is not feasible and a stochastic optimization approach seems more appropriate.

Genetic and particle swarm stochastic algorithms are widely used in optimization of electromagnetic problems comprising large number of variables and multidimensional unknown search spaces [31]. As the fabrication resolution of Si disks is in the range of few nanometers and the upper and lower boundaries of the required radii are well defined, the optimization problem is constrained and can be described by a discrete set of values. Such optimization problem is more suitable for Genetic algorithm.

In order to optimize 1$^{st}$ lobe diffraction efficiency we employed the genetic optimization algorithm [32] with three commonly used operators: selection, crossover and mutation. A comprehensive description of the algorithm and the way it was employed is given in appendix A. The parameters for the optimization were the radii of the disks in the super cell and the thickness of the Si layer. The solid blue lines in Figs. 2(a)-2(c) indicate the performance of the optimized deflectors. The substantial improvement in the efficiency is evident, particularly for the wider angles deflectors where the optimized efficiency is larger by more than a factor of two than that of the conventional design approaches. The parameters and obtained efficiencies of the various design approaches are summarized in Table 1.

Table 1. Super cell radii [nm] obtained from simulations/optimization and deflection efficiency [%] near $\lambda_0$=1550nm.

|  |  | R1 | R2 | R3 | R4 | R5 | Eff. |
|---|---|---|---|---|---|---|---|
| 42.6° | Opt. | 220 | 260 | 316 | - | - | 54 |
|  | Phase | 247.3 | 285 | 305.1 | - | - | 22.3 |
|  | Linear | 236.8 | 285 | 333.2 | - | - | 10 |
| 30.5° | Opt. | 222 | 255 | 275 | 319 | - | 57.5 |
|  | Phase | 237.8 | 276.3 | 290.7 | 311.3 | - | 38.2 |
|  | Linear | 230.8 | 267 | 303 | 339.2 | - | 16 |
| 24° | Opt. | 218 | 252 | 273 | 286 | 323 | 63.4 |
|  | Phase | 230.9 | 268.2 | 285 | 294.2 | 316.2 | 46.1 |
|  | Linear | 227.2 | 256.1 | 285 | 313.9 | 342.8 | 22.6 |

*2.2 Antireflective layer application*

It should be noted that the optimization described above maximizes the transmission efficiency to the upper clad of the device (in our case – SU-8). Although this was the case in most of the previously published studies [3, 4, 16, 17, 21], it is clear that in most practical scenarios the beam emerging from the device will propagate in air (as in [13]). Therefore, it is necessary to add an anti-reflection (AR) layer on top of the upper clad in order to eliminate back-reflection towards the metasurface. A single-layer AR coating should be quarter wavelength thick and have a refractive index value given by the geometric mean of the two surrounding indices: $n_{AR}$=1.257. As such optically transparent material does not necessarily exist; we choose to utilize a commercially available material – amorphous fluoropolymer (CYTOP) as an AR layer. CYTOP has a refractive index of $n_{AR}$=1.334 and relatively negligible absorption at 1550nm [33].

Figure 3 depicts the free-space transmission diffraction efficiencies (TDEs) of the three optimized deflectors as a function of the AR (CYTOP) and the upper cladding (SU-8) thicknesses. Note that the transmission efficiency does not follow the quarter-wave "rule" but rather depends on the thicknesses of the two layers. For the 42.6° deflector, maximal transmission efficiency of 60% is obtained for $d_3$=630nm thick SU-8 layer and $d_4$=980nm thick CYTOP coating. For the 30.5° and 24° deflectors the corresponding values are: 82% efficiency with $d_3$=368nm, $d_4$=576nm and 78% efficiency with $d_3$=837nm, $d_4$=1160nm.

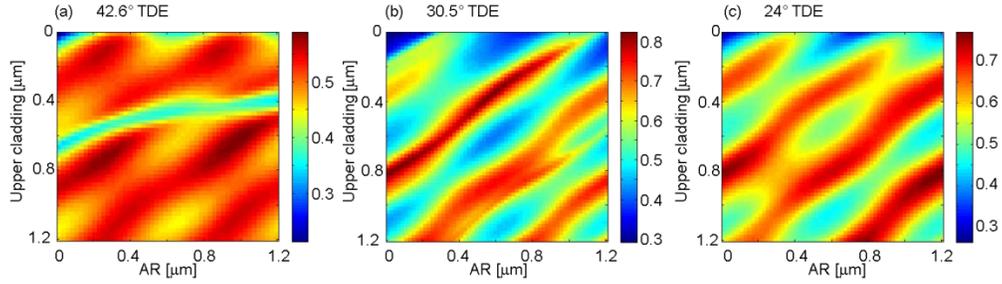

**Fig. 3.** TDE at $\lambda_0$=1550nm with applied AR coating as a function of upper cladding and AR thicknesses of deflectors designed to **(a)** 42.6°; **(b)** 30.5°; **(c)** 24°.

The dependence of the transmission on the thicknesses of the upper clad and the AR layer can be readily understood by examining the transmission properties of a 2-layer cavity as shown in Fig. 4. The upper clad and AR layers form a cavity (Fig. 4(a)) whose transmission properties ($|E_{out}/E_{in}|^2$) are modulated by the thicknesses of the layers. Maximal transmission is obtained when the roundtrip phase (taking into account the propagation angle) in each layer is a multiple integer of $2\pi$. Figure 4(b) depicts the transmission properties of such cavity for the 30.5° deflector. The good agreement with Fig. 3(b) is evident.

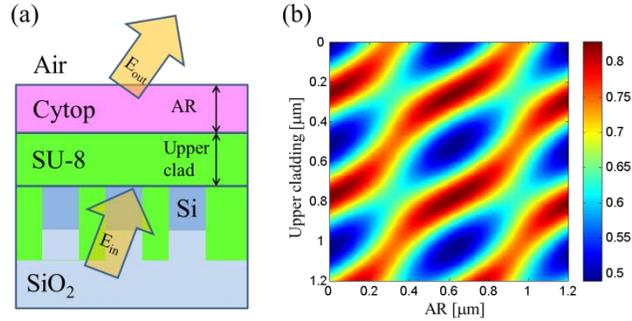

**Fig. 4.** (a) 2-layer cavity model; (b) power transmission through a 2-layer cavity for the 30.5° deflector case.

It is notable that the addition of the AR layer enhances the transmission efficiency of the deflector even further, exceeding 80% for the 30.5° deflector. Figure 5 depicts the spectral response of the transmittance and reflectance diffraction efficiencies (RDE) of the 30.5° deflector, with (b, d) and without (a, c) the AR layer. By comparing two scenarios we can identify the origin of the substantial enhancement of the efficiency due to the AR layer. As can be seen in the figure, the addition of AR layer strongly suppress both the backscattering and higher order transmission lobes in the vicinity of the target wavelength (1550nm), thus yielding high deflection efficiency (1$^{st}$ lobe). We attribute the suppression to the 'Fabry-Perot' like resonances generated by the clad and AR layers. The reflected waves from the clad-AR interface re-excite the dielectric antennas, resulting in destructive interference of the undesired diffraction orders.

For many applications, particularly telecommunications, not only the maximal deflection efficiency is important but also the spectral bandwidth over which the deflector operates efficiently. The 90% bandwidth (spectral range at which the deflection efficiency is at least 90% of maximum value) are17nm, 10nm and 27nm for 42.6°, 30.5° and the 24° deflectors respectively.

Figure 5(e) depicts the calculated phase-front ($E_y$) obtained using FDTD simulations for the 30.5° deflector. The free space propagation angle found by the simulation validates the optimized designs. Note that almost ideal phase-fronts are obtained due to the high transmission efficiency into the 1$^{st}$ diffraction order (i.e. to the desired direction). As mentioned in the beginning of Section 2, the deflector can generate higher order Bragg lobes because its periodicity (the length of the supercell) is larger than the wavelength. However, these diffraction orders can be suppressed over a limited bandwidth, as shown in Fig. 5, where the magnetic and electric dipole resonances of the nanostructure overlap.

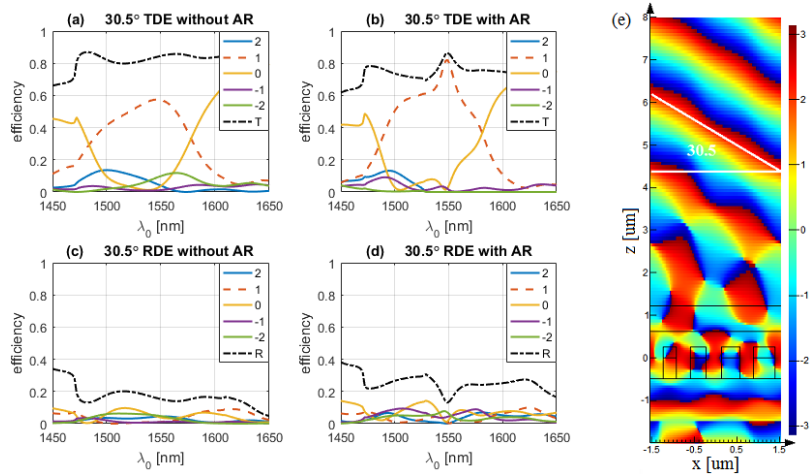

**Fig. 5.** TDE/RDE into particular order (marked by number), total transmittance (T) and reflectance €with **(b, d)** and without **(a, c)** AR layer for the 30.5˚ deflector. €Snapshot of the phase-front ($E_y$) of the light scattered by the 30.5° deflector.

## 2.3 Genetic optimization effectiveness for different wavelengths and design approaches

To further emphasize the superior performance of GA optimization we compare performances to recently published gradient phase approach design [13] (Hexagonal symmetry and varying TiO$_2$ pillars position on silica substrate, $\lambda_0$=580nm) and effective index binary blazed grating design [12]. Note that due to the different choice of materials and wavelengths, high transmission efficiency can be obtained directly to air and the AR layer discussed in the previous section is not required. For comparison with ref. [13] we choose rectangular grid and perform optimization to TiO$_2$ pillars thickness and radii with minimum radius of 50nm due to fabrication constraints. Figure 6(a) shows significant efficiency improvement, particularly for high deflection angles. For example, 87.2% deflection efficiency for the large deflection angle (65°) can be achieved by the genetic optimization approach, yielding the following parameters: array period - 214nm, TiO$_2$ thickness - 450nm and disk radii of 51nm, 82nm, and 106nm. Note the genetic optimization approach can provide significant enhancement of the performance (efficiency and, in fact, the numerical aperture) of metasurfaces based flat lenses. Compared to Ref. [13], the transmission efficiency at 65° is larger by almost 20%.

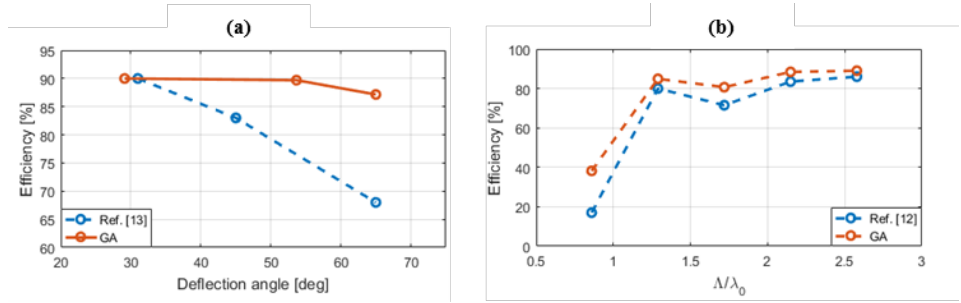

**Fig. 6. (a)** Diffraction efficiencies into the 1$^{st}$ order: GA vs. the design approach of [13]. **(b)** Diffraction efficiencies into the 1$^{st}$ order: GA vs. the design approach of [12].

For comparison with Ref. [12], square $TiO_2$ pillars on silica substrate arranged in rectangular grid were chosen and the wavelength is 633nm. Here as well, the optimization was performed on the $TiO_2$ thickness and dimensions of each element in an array. Considerable diffraction efficiency improvement is obtained across the whole range of array periodicity studied in [12] as evident from Fig. 6(b).

## 3. Conclusion

To conclude, we presented an optimized design approach for flat all-dielectric metasurface devices and utilize it to design highly efficient deflectors for wide angle free space propagation. Our approach is based on utilizing genetic optimization algorithm in order to determine the scatterers dimensions (lateral and vertical) within a unit-cell of the metasurface. Compared to the more commonly used design approaches, our approach provides substantially larger deflection efficiencies, particularly at wide deflection angles (60% at deflection angle of 42.6° at the IR and almost 90% at visible). We attribute the enhanced efficiency to coupling effects between adjacent dielectric scatterers which are often effectively neglected, but are taken into consideration in our design approach.

We also introduced an additional layer on top of the upper cladding which serves as impedance matching layer to air. We found that proper choice of the thicknesses of the upper cladding and the AR layers can enhance the transmission performances of the metasurface substantially, reaching 80% at some angles at the IR spectral range. We attribute this enhancement to resonating interference effects between the dielectric layers. Although the design method presented here was employed for optimizing the performances of metasurfaces based beam deflectors, it is not limited to this application. It should be emphasized that based on the deflector optimization it is possible to design more complex phase masks which realize a variety of highly-efficient flat optical devices such as lenses, holograms, etc. Metasurfaces designed according to our optimized approach are expected to exhibit comparable (and even superior in some cases) performances to those obtained by geometrical phase design (PB) for *linearly* polarized and possibly for unpolarised light. Such component can be highly attractive for various applications such as photonic integrated circuits, Li-Fi applications, displays, and holography.

**ACKNOWLEGMENTS**

The authors thank the XIN Center for partially supporting this research.

**Appendix A: Genetic Algorithm**

Genetic optimization [31] is a stochastic algorithm which aims at optimizing a set of parameters in order to maximize (or minimize) a target function. In our case, the parameters to be optimized are the radii and thickness of the nanostructures comprising the metasurface and the target function is the scattering efficiency to the desired direction. The optimization procedure starts by randomly selecting an initial population of $N$ samples (i.e. $N$ sets of radii and thicknesses) that constitute potential metasurface designs. The values of the initial parameters of each potential design are then transformed to a binary representation and concatenated to form string of bits. Each string forms a "chromosome" – one possible solution to the problem (i.e. a metasurface design). Note that the number of bits which are assigned to each parameter determine the design resolution as well as its upper and lower bounds. Therefore the number of bits assigned to each parameter should reflect the fabrication constraints such as resolution, dimensions, layers thicknesses, etc. After the initial population of chromosomes is created, the GA proceeds to generate a new generation of chromosomes which is derived from the previous one with the aid of three commonly used operators: selection, crossover and mutation.

The selection operator represents the principle of "survival of the fittest". First, the target function is calculated for each of the initial N chromosomes. In our case, the electromagnetic scattering problem is solved numerically for the metasurfaces corresponding to the chromosome and the resulting diffraction efficiency used as fitness function for the next step in the algorithm. A new generation (children) derived from the existing one (parents) by using a procedure referred to as the "fortune wheel". Each individual parent chromosome is assigned with a survival probability represented by a section on roulette wheel with an area which is proportional to the fit function (here - the diffraction efficiency). The children generation is obtained by "rotating" the fortune wheel N times in order to randomly choose one of the chromosomes. Note that as the section size assigned to each chromosome is proportional to its "fitness", the selection process ensures, on average, the selection of most fitted parents and does not limit the number of appearances of the same parent.

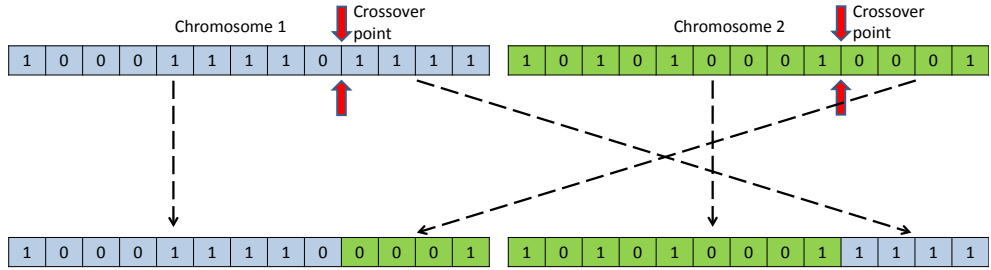

**Fig. 7.** An example for crossover operation

The crossover operator modifies the children generation by randomly mixing couples of chromosomes with probability $p_{cr}$. As illustrated in Fig. 7, the crossing of the two chromosomes is performed in the following manner: the swapping is randomly chosen using uniformly distributed probability. Then, the "left" subarray of chromosome 1 is concatenated with the "right" subarray of chromosome 2 to yield a new chromosome. The "left" subarray with chromosome 2 is concatenated with the "right" subarray of chromosome 2 to yield a second new chromosome. Figure 7 illustrates typical crossover operation. As noted above, the crossing operation is carried out in probability $p_{cr}$. If the crossing is not carried out, the original two chromosomes remain unchanged and moved to the next generation. We note that it is also possible to perform the crossover operation at any sub-section of chromosome (i.e. replacing an internal section between two points). However, we found that in our case such approach has not improved the final efficiency or the number of iterations needed for convergence. The crossover operation is repeated N/2 times in order to obtain *N* new chromosomes.

The mutation operator scans over each chromosome of the newly generated children population (after crossover operator) and randomly flips the bits (0 to 1 or 1 to 0) with probability $p_{mut}$. The *N* chromosomes obtained after this operation serves as the starting point for the next iteration of the process.

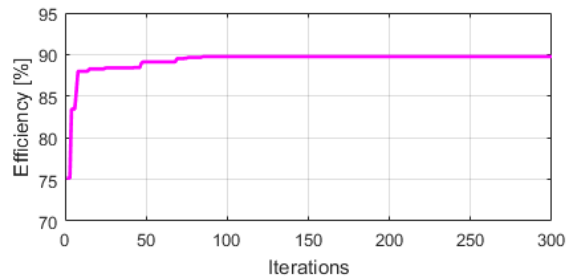

**Fig. 8.** Typical convergence plot for GA.

To summarize, the selection operator choses the best fitted candidates to survive and form the parents of the next generation. The crossover operator facilitates the identification of local maximum in the vicinity of the near best fitted designs and the mutation operator guarantees that a wide region of parameter space is searched in order to find best design. To optimize the metasurface deflection efficiency, we used a population of 20 chromosomes with typical length ranging from 25 to 40 bits. The probability for crossover operation was $p_{cr}$=0.85 and probability for mutation operation was $p_{mut}$=0.05. Figure 8 depicts the convergence process for the design of the 3 disk deflector and chromosome string length of 28 bits. Convergence of the efficiency was typically achieved after 100 iterations.